# Charcoal obtained from cherry stones

A. Venegas-Gomez[1], M. Gomez-Corzo[2], A. Macías-García[3]*, J.P. Carrasco-Amador[4]

*[1]Department of Physics and SUPA, University of Strathclyde, Glasgow G4 0NG, UK*

*[2]Department of Inorganic chemistry. Science Faculty. University of Extremadura, Badajoz, Spain*

*[3]Department of Mechanical, Energetic and Materials Engineering,*

*School of Industrial Engineering, University of Extremadura, Badajoz, Spain*

*[4]Department of Graphical Expression, School of Industrial Engineering,*

*University of Extremadura, Badajoz, Spain*

**Abstract**

The exploitation of using agro-industrial residues as the cherry stone to produce charcoal and activated carbon is very relevant nowadays due to the high demand of these materials and their environmental advantages.

This work describes the methods to prepare charcoal from cherry stones (CS) using final heating temperature of 600 °C, with or without the application of an equal flow of nitrogen or air. The isothermal time has been of 2 hours, an adequate time to study carbonisation by varying the atmosphere. The charcoal obtained has the appropriate characteristics as precursors of activated carbon. Of the three samples prepared in different atmospheres, *N, N~ and AN, the third is an activated carbon.

It is evidenced that the yield in material carbonaceous is somewhat lower when applying the nitrogen current and approximately 5% less when applying the air current. For the different samples, in this work we have studied the effects of the atmosphere during the heat treatment of CS, the yield, the chemical composition and the structure, as well as the pore structure of the same.

*Keywords: charcoal; carbon; carbonisation; filtration; pyrolysis; cherry stone; porosity.*

## 1. Introduction

Agricultural, forestry and industrial residues of lignocellulosic nature (where the main components are cellulose and lignin, as in wood and fruit stones) are suitable for the production of charcoal and activated carbon.

The process of heating organic materials in the absence of air is called pyrolysis or carbonisation. Generally, the term pyrolysis is used when this process focuses on obtaining the gases and oils that are produced, and carbonisation when the process is directed towards the resulting solid product (charcoal) [1].

The carbonisation of the raw material is usually a prior step in the manufacture of the activated carbon, which results from an additional treatment with gases or chemical products in order to increase the porosity of the resulting product [2][3].

The characteristics of the starting material, the conditions of the carbonisation process, and the catalysis process deeply affect the resulting charcoal [4].

The cherry stone is an agro-industrial residue produced in large quantities in the Jerte Valley Cooperative Association when manufacturing cherry or kirsch liquor. The advantage of its utilisation iys a very suitable solution to obtain charcoal and activated carbon due to the current high demand.

In this work, the methods of preparation of charcoal from the cherry stone (CS) are described using different atmospheres. The temperature and time used in carbonisation are suitable for the subsequent production of activated carbon [5] [6]. In Sec. 2 the characterization techniques are described. The results are presented and discussed in Sec. 3 and the conclusions obtained are highlighted in Sec. 4. Additionally, an appendix contains the experimental results in tabular form.

## 2. Experimental procedure

### *2.1. Charcoal preparation*

The system used in the preparation of carbonates is shown schematically in Fig. 1. It consists of a vertical cylindrical furnace connected to a temperature/time programmer; and of a cylindrical stainless-steel reactor, equipped with a threaded cap and two conduits for gas inlet and outlet [7][8].

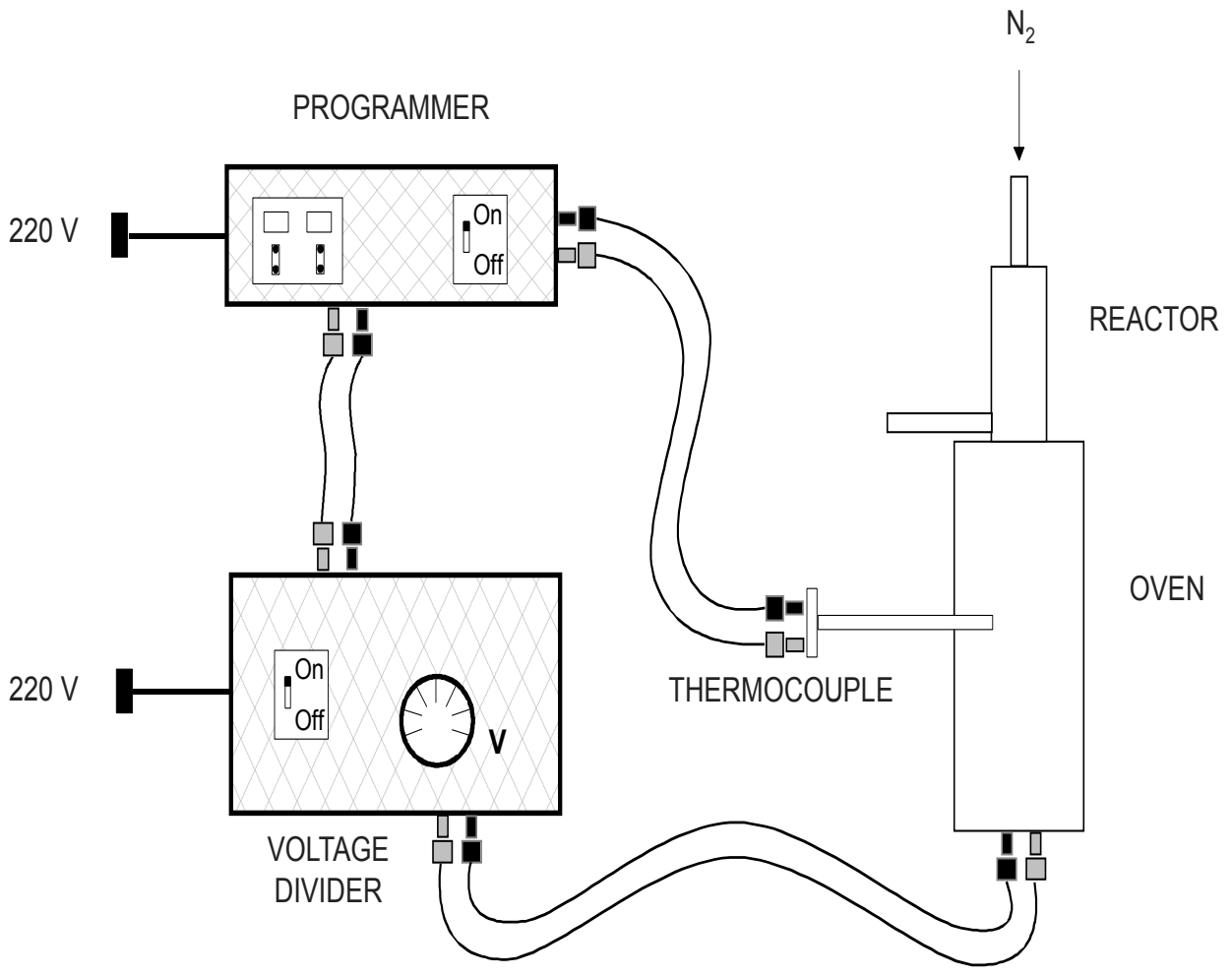

**Fig. 1. Experimental system used in the preparation of samples.**

To study the influence of the heat treatment atmosphere, three samples have been prepared under the following conditions:

- Mass of CS: 100g.

- Carbonisation atmosphere: atmosphere formed during heating, with and without nitrogen (200 $cm^3$ $min^{-1}$) or air flow (200 $cm^3$ $min^{-1}$).

- Initial heating temperature: room temperature.

- Heating rate: 10 °C $min^{-1}$.

- Final heating temperature: 600 °C.

- Isothermal heating time (t) at 600 °C: 2 h.
- Atmosphere during cooling to room temperature: nitrogen (200 $cm^3$ $min^{-1}$).

The samples are represented in Table 1: *N, N~ and AN. In the table is specified, based on the nomenclature indicated, that 600 is the isothermal heating temperature the samples; 2 the number of hours of isothermal heating at 600 °C; N stands for nitrogen; * indicates the atmosphere of heating and cooling it is at the beginning; ~ indicates the atmosphere of cooling it is at the end; and A means air.

**Table 1. Charcoal preparation**

| Sample | Precursor | Heating speed (°C $min^{-1}$) | Flow of gas ($cm^3$ $min^{-1}$) | Isothermal temperature (°C) | Isothermal time (h) |
|---|---|---|---|---|---|
| *N | CS | 10 | 0 | 600 | 2 |
| N~ | CS | 10 | 200 ($N_2$) | 600 | 2 |
| **AN** | CS | 10 | 200 (A) | 600 | 2 |

This rate of heating is frequent in investigations of carbonisation processes [9]. Brunner and Roberts [10] have already found that the carbonisation yield of the cellulose increases with the decrease in the heating rate. Mackay and Roberts [11] also found the same effect when studying the influence of pyrolysis conditions on the yield and microporosity of lignocellulosic carbonates [12]. On the other hand, Kumar et al. [13] have shown that a carbonisation of acacia wood at 1050 °C under rapid carbonisation clearly shows the absence of secondary carbon deposit, whereas the reverse occurs when carbonisation is slow. More recently, Lua and Guo [14] have found that 10 °C min-1 and 150 $cm^3$ $min^{-1}$ are optimum conditions in the pyrolysis of carbonates prepared from residues of extracted palm fibre.

*2.2. Charcoal characterisation*

The charcoal and activated charcoal resulting from the following processes of preparation have been chemically and structurally characterized.

*2.2.1 Chemical characterization*

The immediate chemical composition and the organic chemical structure have been determined [15].

*2.2.2 Pore structure characterization*

Samples have been characterized by applying density (packaging, apparent and real), mercury porosimetry and scanning electron microscopy techniques [16]. In addition, they have been characterized by physisorption ($CO_2$, 273.15 K; $N_2$, 77 K) in a Micromeritics automatic apparatus (model ASAP 2010).

The method followed in the determination of adsorption isotherms is recorded in the literature [17] [18]. Approximately 0.5g were degassed at 250 °C for 24h under a dynamic vacuum of $10^{-4}$ mm Hg (1 mm Hg = 1.3333 $10^{2}$ Pa). After cooling, the samples adsorbed carbon dioxide to 273.15 K or nitrogen at 77 K, under different pressures.

The application of the DR [15], DA [16], and S [16] equations to the adsorption data has provided the micropore volumes of samples accessible to carbon dioxide, $W_{DR}(CO_2)$, $W_{DA}(CO_2)$ y $W_S(CO_2)$, and the values of the volume of micropores accessible to nitrogen, $W_{DR}(N_2)$, The values of the gas volume conversion factor in liquid volumes at the adsorption temperatures were 1.831 $10^{-3}$ (density 1.08 g $cm^{-3}$) [21] for carbon dioxide and 1.547 $10^{-3}$ (density 0.808 g $cm^{-3}$ [22] for nitrogen. The values taken for the affinity coefficient were 0.48 for carbon dioxide [23] and 0.34 for nitrogen [20].

The application of the DR equation has also provided the values for the parameter $E_o$, with which the mean values for the size of slit-shaped micropores ($L_o$) have been calculated, by applying the semi-empirical correlation of Stoeckli and Ballerini [24].

The microporosity distributions have been calculated by the function proposed by Stoeckli [20] applied to the adsorption results of carbon dioxide at 273.15 K.
From the nitrogen adsorption data, the samples specific surface area $S_{BET}$ [17] was determined following the *IUPAC* recommendations: linearity range ($p/p^o<0.30$) and mean area of the $N_2$ molecule occupied in the monolayer 0.162 $nm^2$ [25]. From the carbon dioxide adsorption data, the specific micropore surface area of the samples $S_{mi}$ has been determined [20]. The total surface area of the samples, S ($CO_2$) [26] with the values of $W_{DR}$ ($CO_2$) and $W_{DA}$($CO_2$), and considering the mean area of the $CO_2$ molecule as 0.187 $nm^2$ [27].

## 3. Results and discussion

Of the three samples prepared in different atmospheres (*N, N~ and AN), the third is an activated carbon. In the following sections, to better understand the effects of the atmosphere during the heat treatment of CS on the yield, the chemical composition and structure, as well as the pore structure of the samples, have been included.

### *3.1. Yield*

As can be seen in Table 2, the yield is somewhat higher for *N compared to N~ , and significantly lower for AN. Considering that the volatile matter produced in the heat treatment of CS passes through the pores of the carbonaceous material that is formed to its surface, the application of a nitrogen stream, in addition to providing an inert atmosphere for the pyrolysis, help to eliminate the volatile matter, forming a smaller amount of secondary carbon in the walls of the pores of N~. When applying an air stream, the oxygen in it acts as a reagent, whereby the deposit of volatile matter in the pores of AN is significantly smaller. In Table 2 the yield is with respect to 100 g of original sample (CS) on dry basis (CS without moisture). That is why in CS there is nothing.

**Table 2. Yield and immediate analysis of raw material and samples *N, N~, and AN.**

| Sample | Yield | Original sample reference (%) | | | | Dry sample reference (%) | | |
|---|---|---|---|---|---|---|---|---|
| | (%) | Humidity | Volatile matter | Ashes | Fixed carbon | Volatile matter | Ashes | Fixed carbon |
| **CS** | - | 5.41 | 71.72 | 0.24 | 22.63 | 75.82 | 0.25 | 23.93 |
| ***N** | 26.32 | 6.94 | 7.75 | 1.36 | 83.95 | 8.33 | 1.46 | 90.21 |
| N~ | 25.36 | 6.87 | 6.42 | 1.49 | 85.22 | 6.89 | 1.60 | 91.51 |
| **AN** | 21.43 | 7.12 | 5.94 | 1.31 | 85.63 | 6.40 | 1.41 | 92.19 |

The quality and performance are similar to those obtained using other lignocellulosic materials (wood, peel and fruit stones). It was especially compared with the treatment applied to the rockrose on a dry basis (free of moisture).

*3.2. Chemical composition and structure*

According to the previous comments, the immediate analyses (Table 2) show that decreasing the yield also decreases the volatile matter. The low ash content and high fixed carbon content of the three samples are favourable properties to be used as activated carbon precursors.

According to the FTIR spectra (Fig. 2) the thermal treatments of CS have produced strong changes in the chemical-organic structure of the material. In the spectra of *N, N~, and AN, a wide band at wavelengths less than 1600 $cm^{-1}$ can be observed, which is the result of the overlap of several individual absorption bands. The resulting samples contain aromatic rings (at 1550 $cm^{-1}$) with the highest amount of aromatic C-H bonds for *N and AN (bands better defined at 900-700 $cm^{-1}$). Oxygen forms C-O bonds (1300-1000 $cm^{-1}$ band), without the air atmosphere leading to a more oxygenated structure in the case of AN; at 1240 $cm^{-1}$ could correspond to asymmetric deformation vibrations in C-O-C bonds in a single layer, and the band at 1110 $cm^{-1}$ to C-O-C bonds between graphite layers.

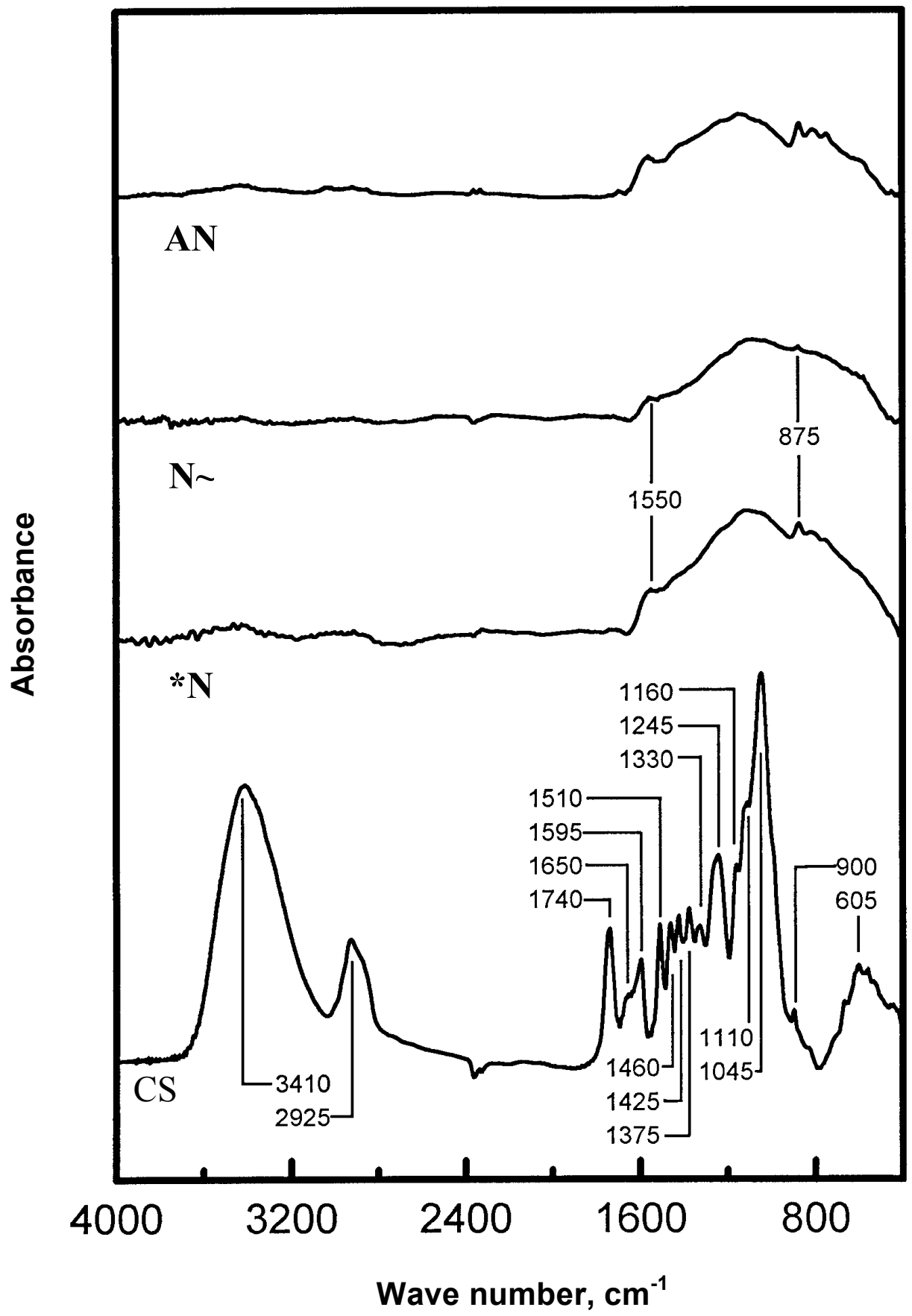

**Fig. 2. FTIR spectra of CS, *N, N~, and AN.**

*3.3. Pores structure*

The values of the density and the total volume of pores accessible to helium at room temperature are given in Table 3. In contrast to CS, the density values (packing density ρ, mercury accessible density $\rho_{Hg}$ and helium accessible density $\rho_{He}$) indicate that the three heat treatments have developed porosity. The $V_{He}$ values show that at least the pore volume has doubled. The highest pore volume of AN is due to the reactive oxygen atmosphere.

**Table 3**. Raw material and samples *N, N~, and AN: densities and total pore volume.

| **Sample** | $\rho$ (g cm$^{-3}$) | $\rho_{Hg}$ (g cm$^{-3}$) | $\rho_{He}$ (g cm$^{-3}$) | $V_{He}$ (cm$^3$ g$^{-1}$) |
|---|---|---|---|---|
| **CS** | 0.76 | 1.09 | 1.38 | 0.193 |
| *N | 0.62 | 1.02 | 1.76 | 0.412 |
| N~ | 0.59 | 1.00 | 1.71 | 0.415 |
| AN | 0.59 | 0.91 | 1.72 | 0.518 |

Table A1 to Table A6 of the appendix give the data on the adsorption of carbon dioxide at 273.15 K and nitrogen at 77 K. With them, the isotherms of Fig. 3 and Fig. 4 have been obtained, showing that the treatment atmosphere has a greater influence on nitrogen adsorbent capacity than on the adsorption capacity of carbon dioxide.

According to the classification of the isotherms of physisorption of the IUPAC [21], the isotherms of Fig. 4 can be considered of type I, characteristics of microporous solids. More precisely, the type $I_b$ [28] isotherm can be considered, especially the *N. As nitrogen adsorption grows relatively little when $p/p^o$ increases, the samples have a small amount of mesopores [29]. Therefore, the positions and shapes of the nitrogen adsorption isotherms indicate that the development of the broad microporosity of the samples follows the sequence AN> N~>>*N, and that the development of mesoporosity is somewhat greater in AN and N~.

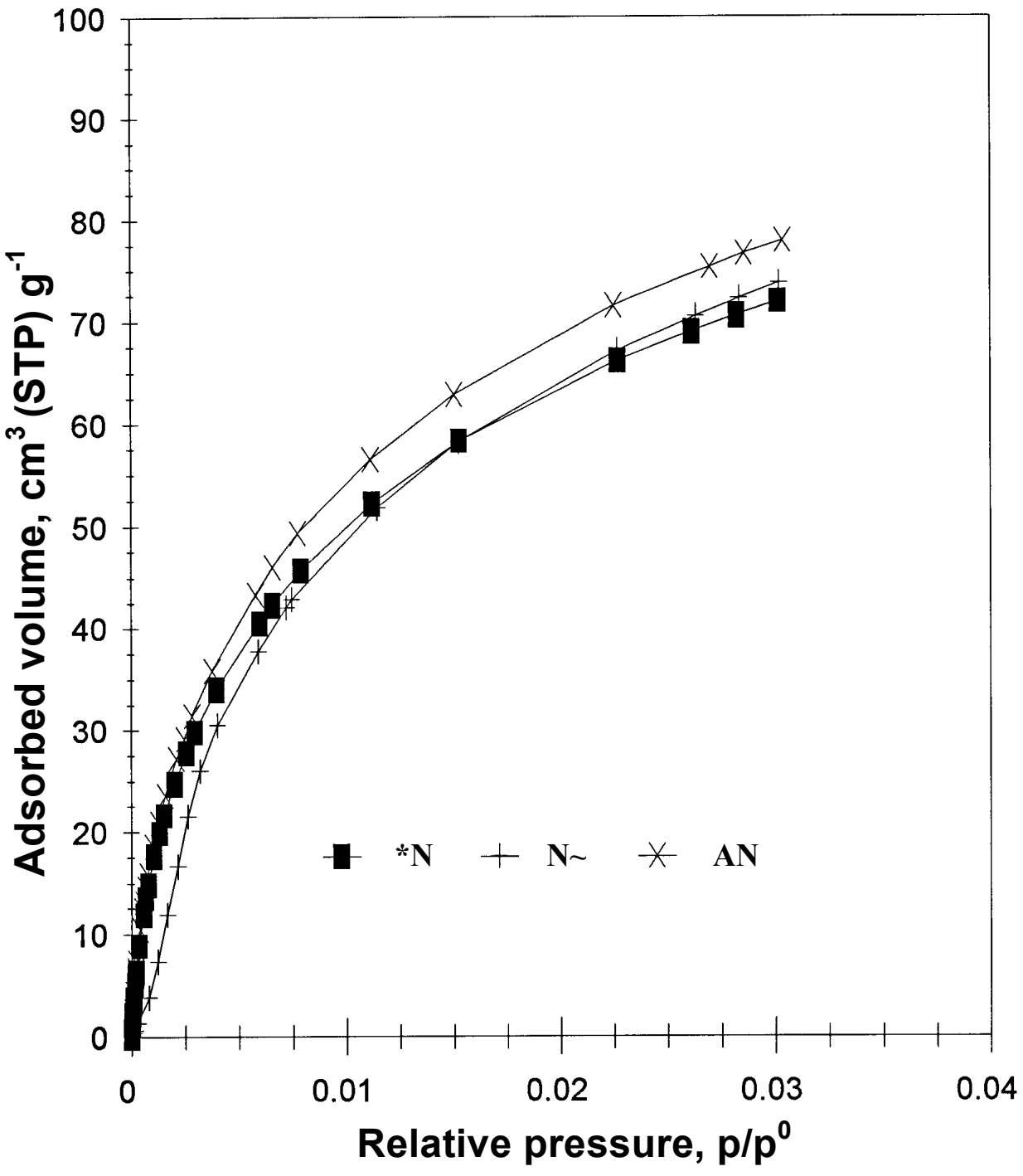

**Fig. 3. Carbon dioxide adsorption isotherms at 273.15 K of *N, N~ and AN.**

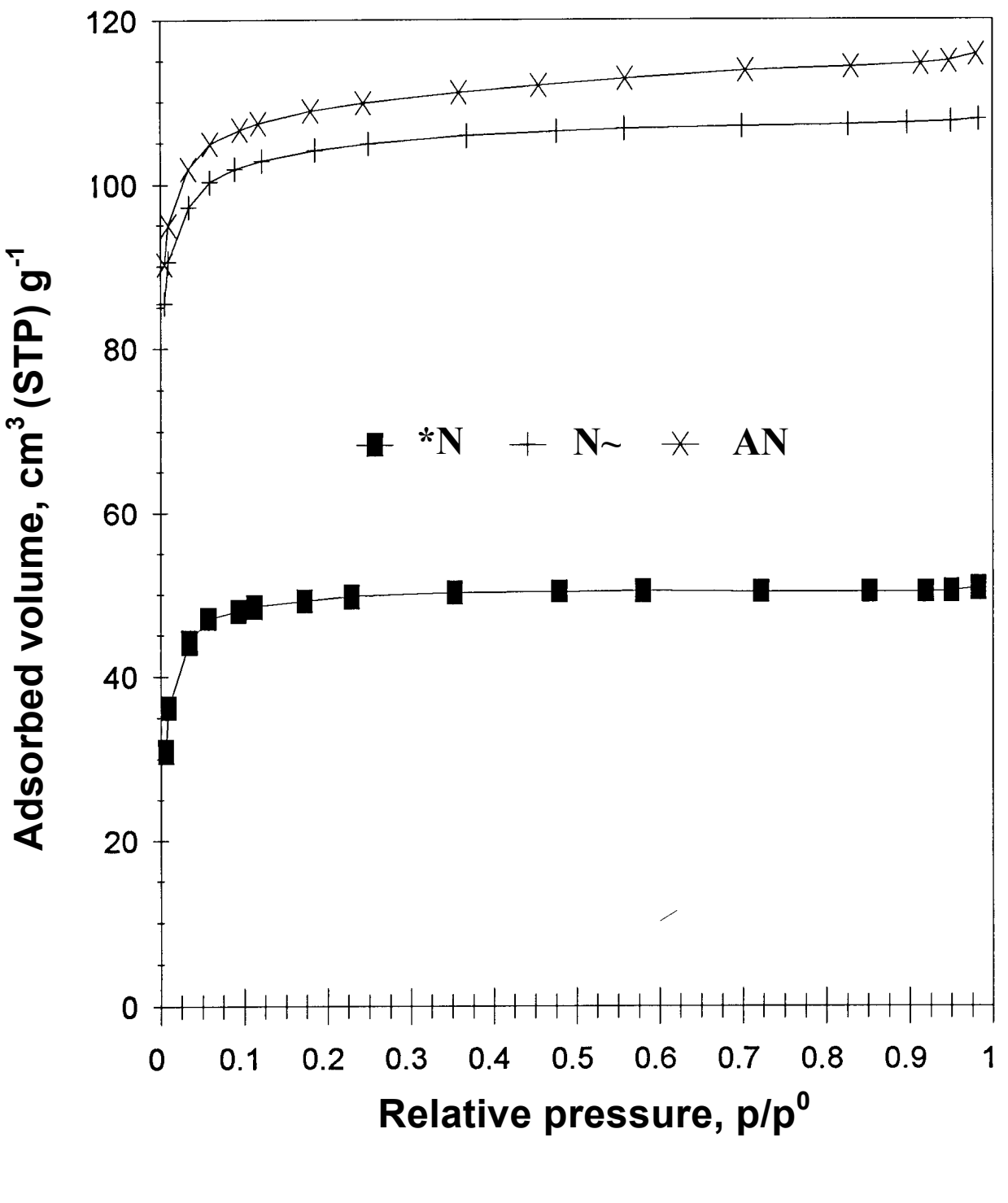

**Fig. 4. Nitrogen adsorption isotherms at 77 K of *N, N~ and AN.**

The calculations made by applying the equations DR, DA and S to the isotherms of Fig. 3 and Fig. 4 are also given in Table A1 to Table A6.

The adjustments of these equations are shown in Fig. 5 to Fig. 8. The DR (CO2) setting of *N (Fig. 5) is of type A of those indicated by Marsh [26], while the DR (N2) setting (Fig. 8) can be considered as type D [26]. Therefore, according to the results found by Rodríguez-Reinoso et al. [30] in characterizing almond shell charcoal, the nitrogen adsorption at 77 K is kinetically restricted at *N, while adsorption of carbon dioxide at 273.15 K is not. The rest of the adjustments can be considered of type A [26].

The values for the parameters in equations DR, DA and S (Table 4 and Table 5) allow to establish that $W_S(CO_2) < W_{DR}(CO_2)$ y que $W_S(CO_2) < W_{DA}(CO_2)$ for each sample. Therefore, the volume of narrow micropores of each carbon depends on the applied equation. As $W_{DR}(CO_2) > W_{DR}(N_2)$ the microporous structure of the three samples consists mainly of narrow micropores. AN is the sample with the largest volume of narrow and wide micropores.

**Table 4. Parameters of the DR and DA equations applied to the adsorption isotherms of samples *N, N~ and AN.**

| | DR Equation | | | | DA Equation | | |
|---|---|---|---|---|---|---|---|
| **Sample** | $W_{DR}(CO_2)$ ($cm^3\ g^{-1}$) | $E_0(CO_2)$ ($kJ\ mol^{-1}$) | $W_{DR}(N_2)$ ($cm^3\ g^{-1}$) | $E_0(N_2)$ ($kJ\ mol^{-1}$) | $W_{DA}(CO_2)$ ($cm^3\ g^{-1}$) | $E_0(CO_2)$ ($kJ\ mol^{-1}$) | $n(CO_2)$ |
| ***N** | 0.214 | 23.43 | 0.078 | 21.06 | 0.239 | 22.31 | 1.91 |
| **N~** | 0.253 | 22.67 | 0.166 | 21.03 | 0.212 | 24.51 | 2.16 |
| **AN** | 0.211 | 24.14 | 0.175 | 20.86 | 0.305 | 20.45 | 1.73 |

**Table 5. Parameters of the S equation applied to the adsorption isotherms of carbon dioxide at 273.15 K of samples *N, N~ and AN.**

| Sample | $K_0$ (nm kJ mol$^{-1}$) | $W_S(CO_2)$ (cm$^{-3}$ g$^{-1}$) | $\nu$ | $a$ (nm$^{-1}$) |
|---|---|---|---|---|
| ***N** | 22.08 | 0.158 | 5.35 | 9.22 |
| **N~** | 20.19 | 0.158 | 7.47 | 18.73 |
| **AN** | 24.41 | 0.177 | 4.27 | 5.02 |

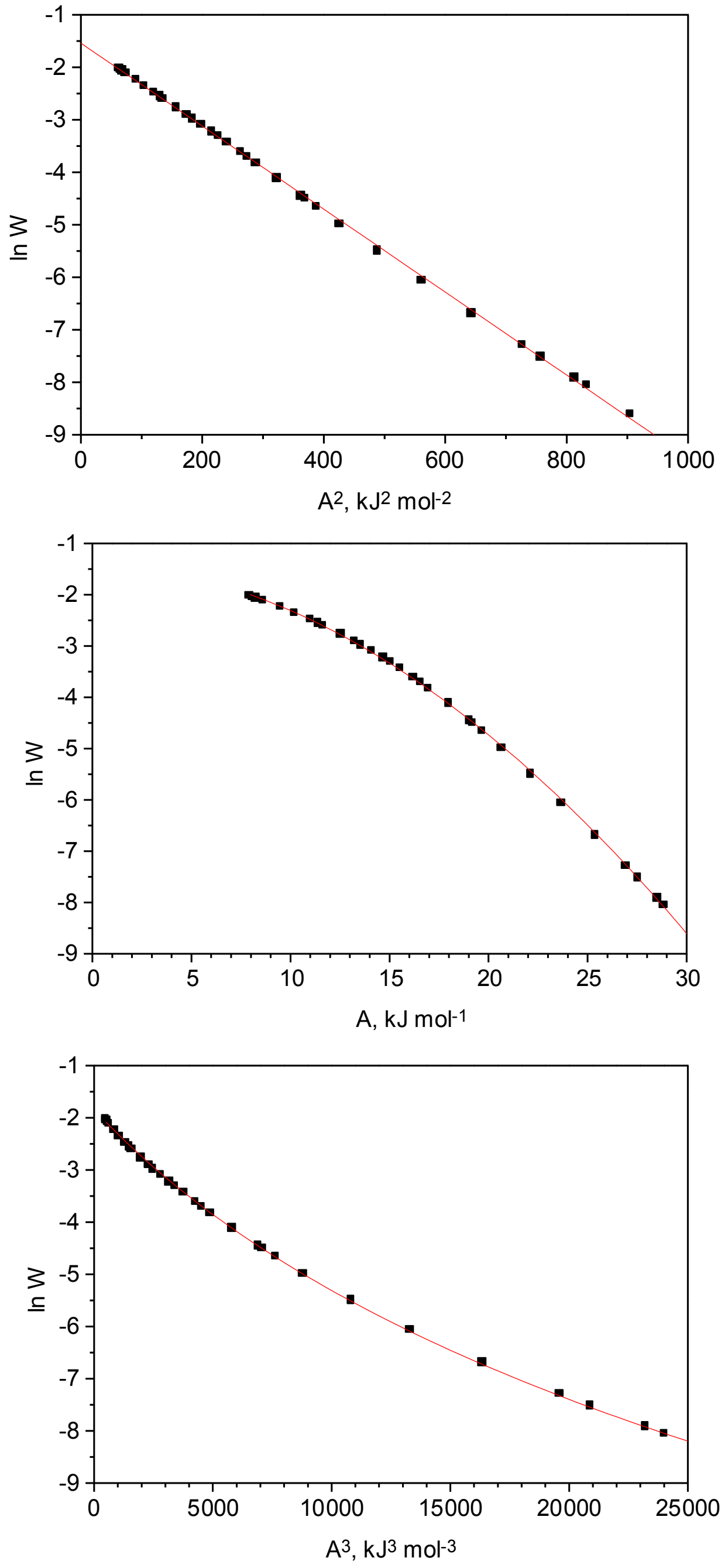

**Fig. 5. Sample *N: application of the DR (above), DA (medium) and S (below) equations to the adsorption isotherm of carbon dioxide at 273.15 K.**

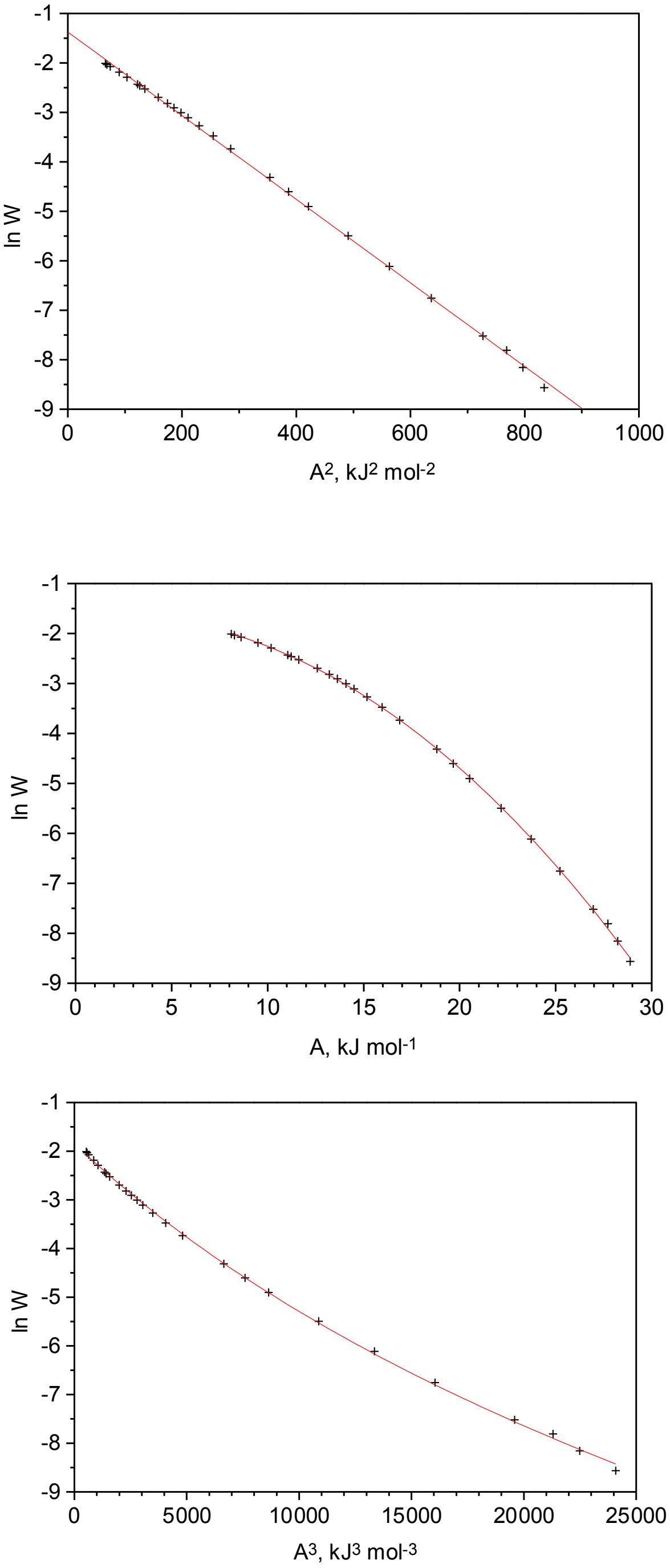

**Fig. 6. Sample N~: application of the DR (above), DA (medium) and S (below) equations to the adsorption isotherm of carbon dioxide at 273.15 K.**

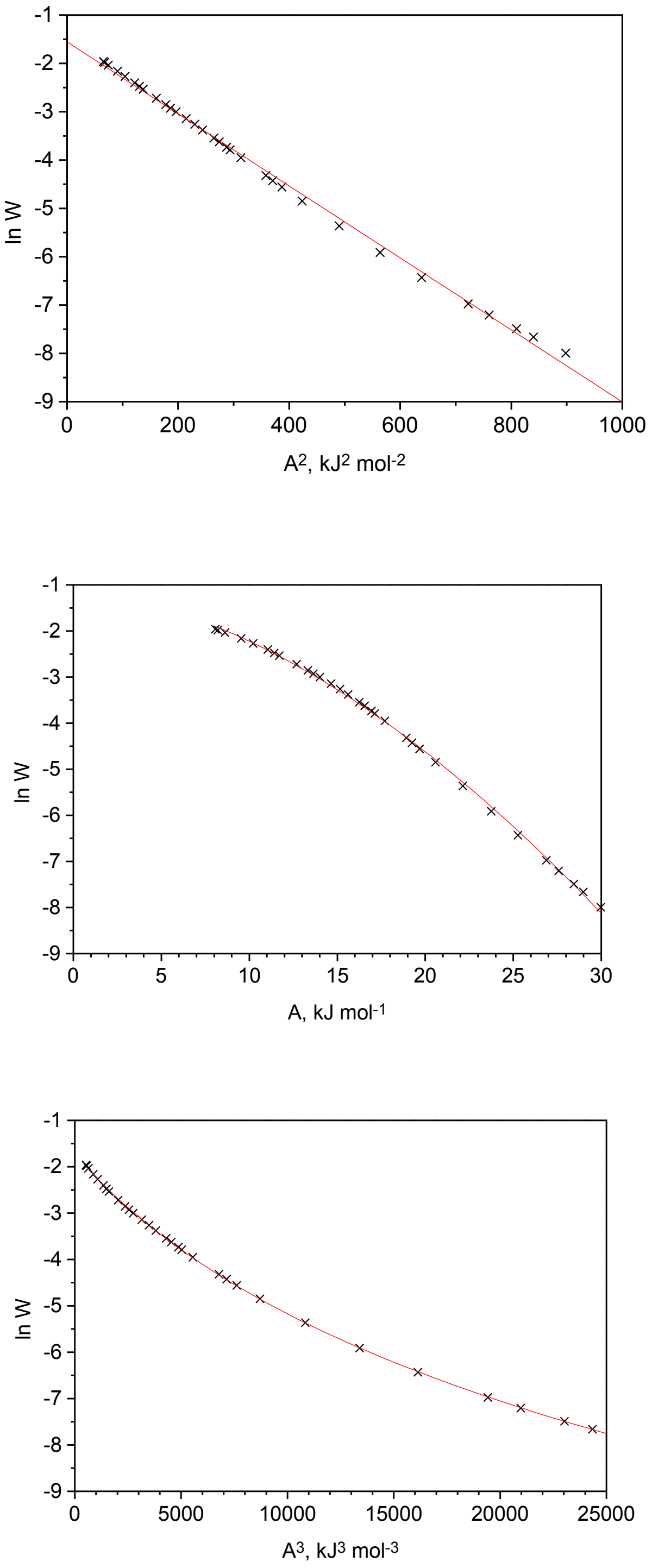

**Fig. 7. Sample AN: application of the DR (above), DA (medium) and S (below) equations to the adsorption isotherm of carbon dioxide at 273.15 K.**

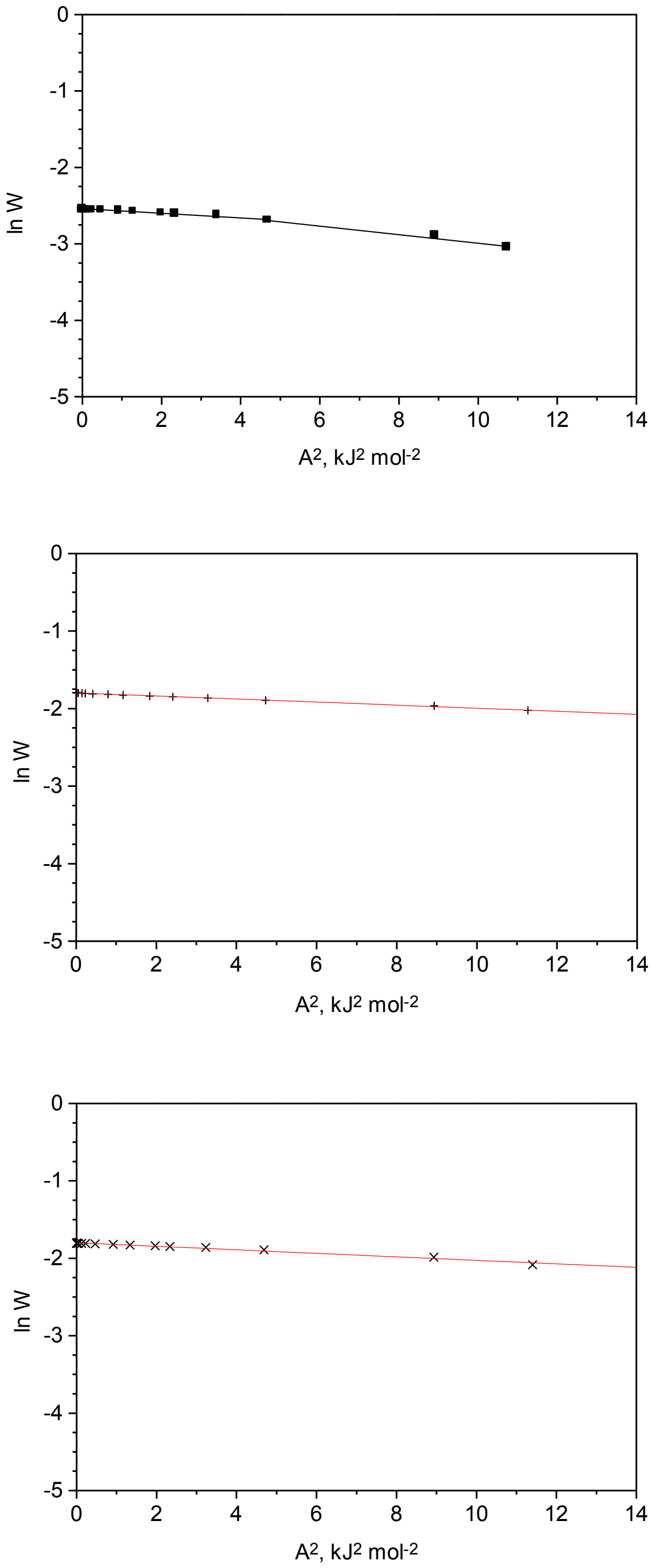

**Fig. 8. Samples *N (top), N~ (medium) and AN (below): application of the DR equation to nitrogen adsorption isotherms at 77 K.**

All the values of the characteristic energy of the DR equation (Table 4) are included in the range of characteristic energies 30-35 to 17-18 kJ mol$^{-1}$ [16], resulting in $E_0(CO_2)$ > $E_0(N_2)$. The values of $K_0$ (Table 5) are between 16 and 35 nm kJ mol$^{-1}$ [31].

The fact that the value of n($CO_2$) for AN is the smallest of the three (Table 4) indicates that the narrow micropore structure of this sample is the most heterogeneous. This can be observed in the representations of Fig. 9, which have the same appearance, differing in the width of the curves and in the maximum value of dW/dL.

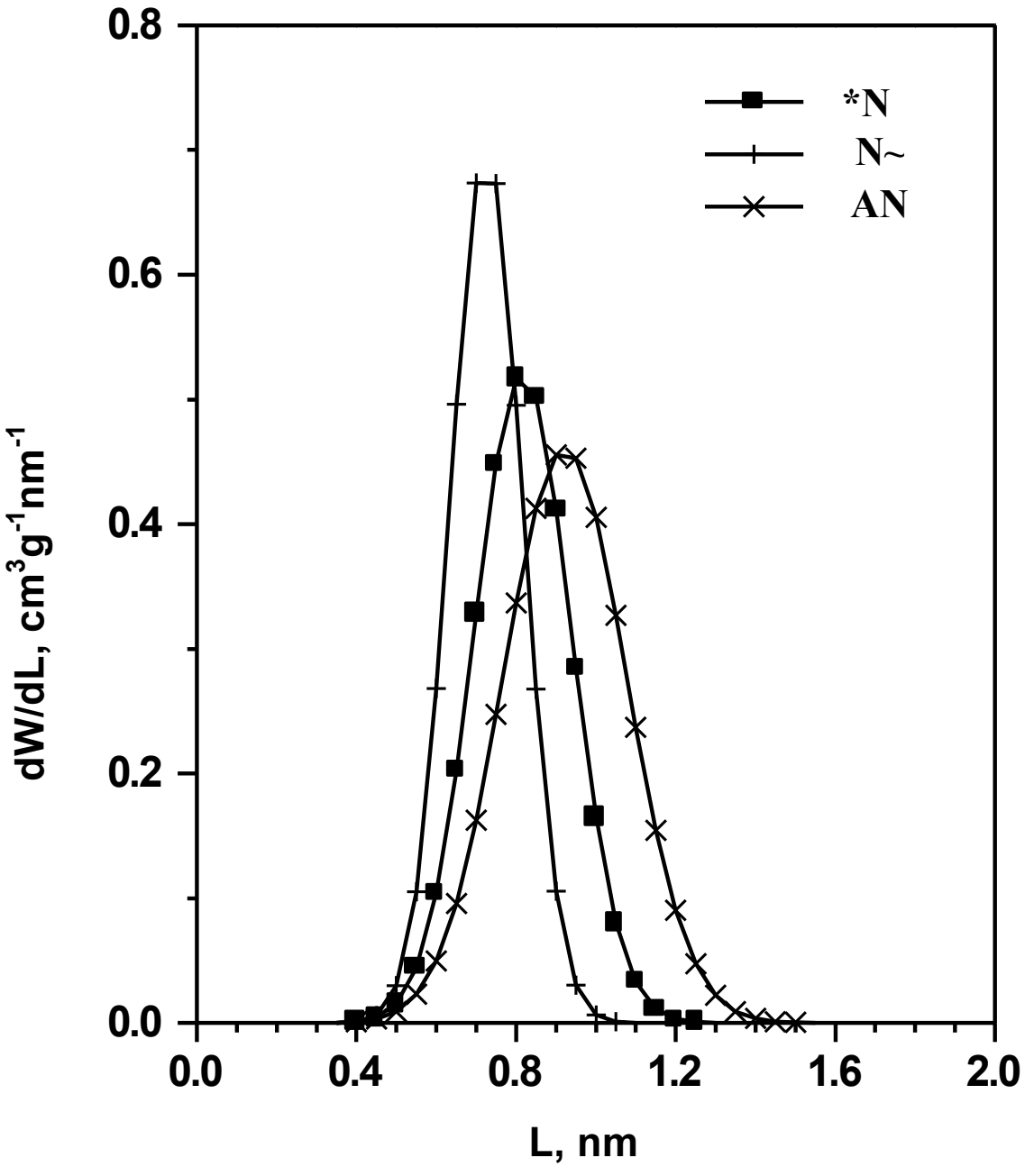

**Fig. 9. Distributions of micropore sizes determined by applying the Stoeckli method to the adsorption isotherms of carbon dioxide at 273.15 K of samples *N, N~ and AN.**

With values of $E_0(CO_2)$ and $E_0(N_2)$, the mean micropore widths $L_0(CO_2)$ y $L_0(N_2)$ have been calculated (Table 3.5), which are included in the range 0.4-2 nm [31].

Further information on the microporous structure of the samples is derived from the values of the surface area (Table 6). The calculations made when applying the BET($N_2$) equation are given in Table A4 to Table A6 and the BET($N_2$) settings are shown in Fig. 10; with the highest value of $S_{BET}(N_2)$ corresponding to AN. From the Stoeckli-Ballerini relation [31], the values of the area of the walls of micropores such as $S_{mi}(CO_2)$ > $S_{mi}(N_2)$ have been obtained for each sample. $S_{BET}(N_2)$ > $S_{mi}(N_2)$ for each sample because $S_{BET}(N_2)$ represents the equivalent area of the monolayer, which would coincide with $S_{mi}(N_2)$ if the micropores accommodated two adsorbate layers. For the same reason, $S_{DR}(CO_2)$ > $S_{mi}(CO_2)$ y $S_{DA}(CO_2)$ > $S_{mi}(CO_2)$.

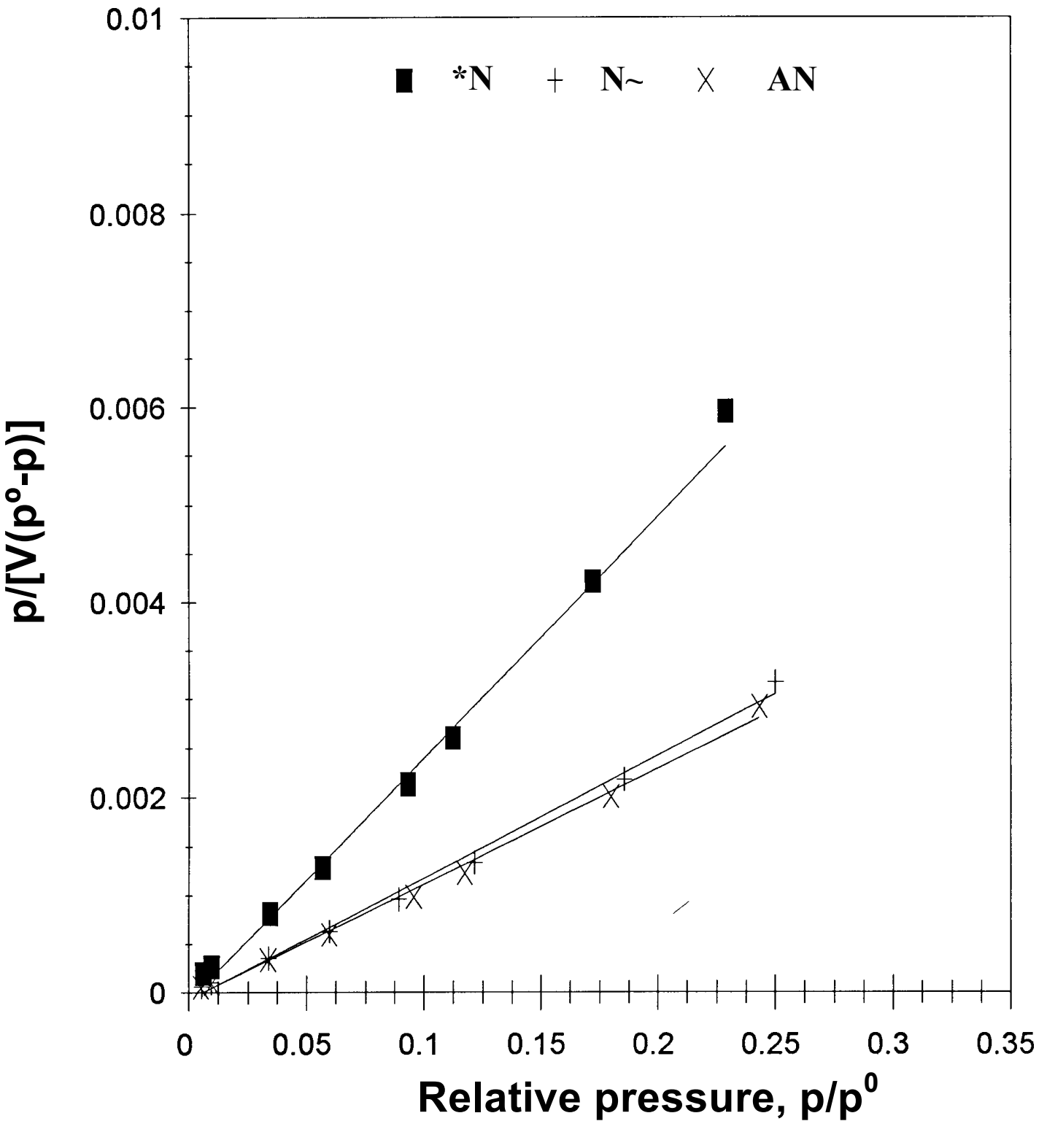

**Fig. 10. BET settings (N2, 77 K) of samples *N, N~ and AN.**

**Table 6. Average size of micropores and specific surface areas of samples *N, N~ and AN.**

| Sample | $S_{BET}(N_2)$ ($m^2\ g^{-1}$) | $L_O(CO_2)$ (nm) | $S_{mi}(CO_2)$ ($m^2\ g^{-1}$) | $L_O(N_2)$ (nm) | $S_{mi}(N_2)$ ($m^2\ g^{-1}$) | $S_{DR}(CO_2)$ ($m^2\ g^{-1}$) | $S_{DA}(CO_2)$ ($m^2\ g^{-1}$) |
|---|---|---|---|---|---|---|---|
| *N | 171 | 0.90 | 476 | 1.12 | 139 | 589 | 558 |
| N~ | 321 | 0.96 | 527 | 1.12 | 296 | 696 | 583 |
| AN | 369 | 0.85 | 496 | 1.14 | 307 | 581 | 839 |

The results obtained in the characterization of the microporous structure of *N and N~, can be interpreted considering that a greater quantity of tarryl matter, deposited and decomposed in the microporous network of 6002N, partially fills or blocks more microporos than in N6002, forming more constrictions. Although both samples are mainly formed by narrow micropores, they have more uniform widths in N~. In AN, in addition to the drag of volatile matter produced in the heat treatment of CS, oxygen reaction of the air stream applied with that matter must occur, preventing its deposit in the carbonaceous material being formed. In comparison with the other two samples, a microporous structure formed mainly of narrow micropores is obtained as well, but in this case is more heterogeneous, more open, and with a larger volume of micropores.

Regarding the non-microporous structure of the samples, it has been commented previously that the isotherms of Fig. 4 indicate that each of them has a small amount of mesopores, since nitrogen adsorption at 77 K grows relatively little when p/p^{o} increases. As can be seen in Table 7, the mesopore volume values ($V_{me}$), determined by mercury porosimetry, are the same (0.006 $cm^3\ g^{-1}$) for all three samples. In contrast, the macropore volume values ($V_{ma}$) are almost the same for *N and N~, and higher in the case of AN. Consequently, the values of $V_{ma}$ and the cumulative volume of mesopores and macropores ($V_{Hg}$) are almost the same, i.e. it turns out that the non-microporous structure of the samples consists mainly of macropores. From the same table, it is also deduced that the three samples have a high volume of macropores compared to CS.

**Table 7. Average size of micropores and specific surface areas of samples *N, N~ and AN.**

| Sample | $V_{ma}$ ($cm^3\ g^{-1}$) | $V_{me}$ ($cm^3\ g^{-1}$) | $V_{Hg}$ ($cm^3\ g^{-1}$) |
|---|---|---|---|
| **CS** | 0.122 | 0.000 | 0.122 |
| ***N** | 0.246 | 0.006 | 0.252 |
| **N~** | 0.248 | 0.006 | 0.254 |
| **AN** | 0.291 | 0.006 | 0.297 |

The representations of Fig. 11 have been obtained with the recorded mercury porosimetry data given in Table A7 to Table A9. The variations of $V_{Hg}$ with the pore radius are practically the same for *N and N~, which shows that the non-microporous structure is not modified by the application of the nitrogen flow during the heat treatment of CS. In contrast, $V_{Hg}$ for AN is greater for pore radii smaller than 400 nm, demonstrating that the air atmosphere produces a greater macroporous development. The representation of the derivative of $V_{Hg}$ versus the pore radius is essentially unimodal, with a peak centred at approximately 200 nm for both the raw material and the three samples prepared.

The results obtained in the characterization of the pore structure make it possible to specify that the effect of the reactive action of oxygen in the air atmosphere mainly results in a better development of the broad microporosity and the macroporosity.

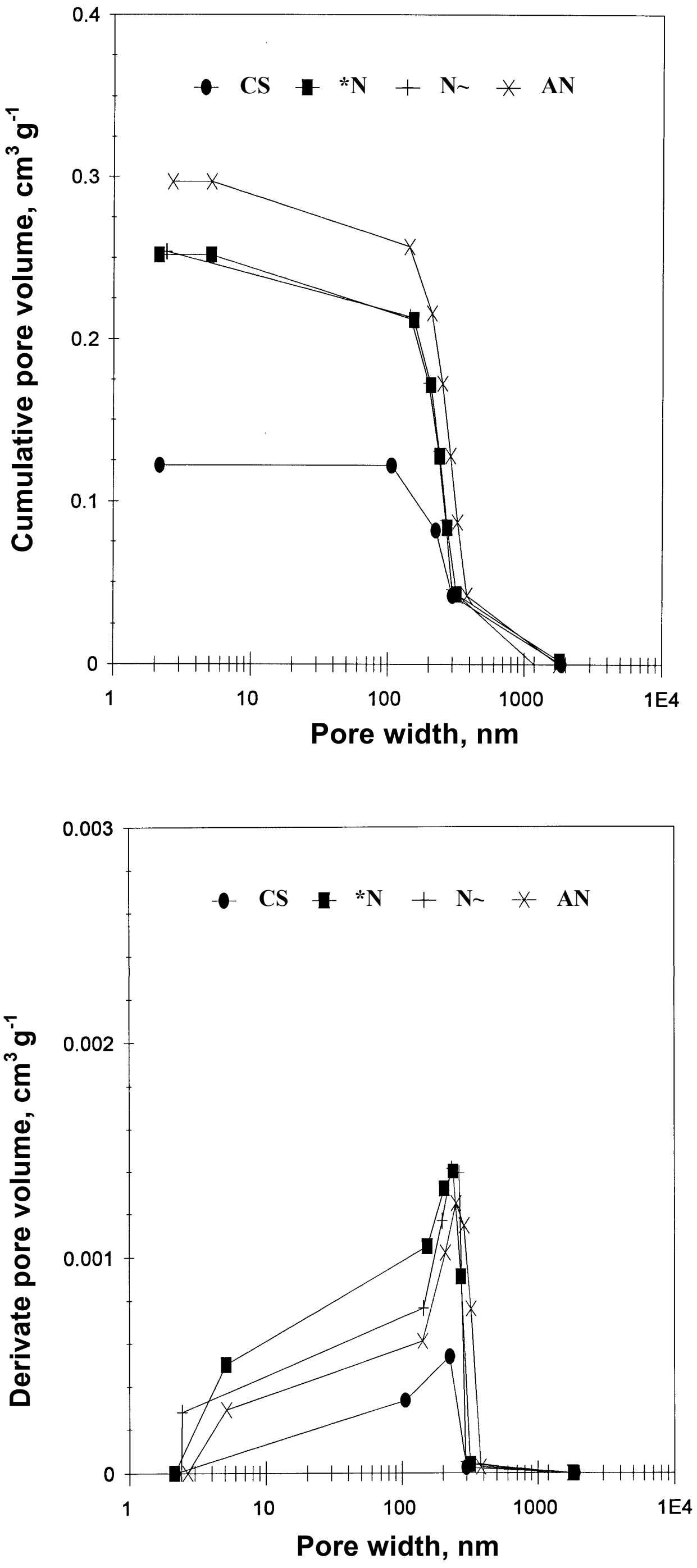

**Fig. 11. Distributions of pore sizes in the mesopore and macropore intervals of samples *N, N~ and AN determined by mercury porosimetry**

## 4. Conclusions

In view of the results presented and discussed in this paper, the following conclusions can be drawn:

- Residual cherry stone treatments of kirsch production carried out at the final heating temperature of 600 °C, with or without the application of an equal flow of nitrogen or air, are three processes of different yield in material carbonaceous material. The yield is somewhat lower by applying the nitrogen stream and about 5% lower by applying the air stream.
- The three prepared samples differ in the immediate chemical composition, having a low ash content, around 1.5%. The oxygen forms ether-like structure, without the air atmosphere leading to a more oxygenated structure.
- Compared with the raw material, the volume of pores accessible to helium at room temperature is at least doubled. The application of the air stream produces a more significant porous development.
- The microporous structures of the samples are mainly formed by narrow micropores, with greater heterogeneity of sizes when air is applied. However, the differences are more significant in wide microporosity; the volume of wide micropores or the BET surface area ($N_2$, 77 K) doubles when nitrogen is applied; and with air improves the development of this microporosity. These results can be interpreted by considering that a greater quantity of tarry matter, deposited and decomposed in the microporous network of the carbonisation prepared without application of nitrogen or air current, partially fills or blocks more micropores, forming more constrictions.
- Regarding the non-microporous structure, the development is important compared to the raw material. The application of the nitrogen stream does not improve the development of the structure of macropores and mesopores, while the air current develops macroporosity. On the other hand, none of them has any influence on the development of mesoporosity.

**Acknowledgements**

The authors appreciate support by the Department of Inorganic Chemistry, Faculty of Science, University of Extremadura.